# Electronic properties and the nature of metal-insulator transition in NdNiO$_3$ prepared at ambient oxygen pressure

M. K. Hooda and C. S. Yadav*
School of Basic Sciences, Indian Institute of Technology Mandi, Mandi -175001, H.P. (India)

We report the electronic properties of the NdNiO$_3$, prepared at the ambient oxygen pressure condition. The metal-insulator transition temperature is observed at 192 K, but the low temperature state is found to be less insulating compared to the NdNiO$_3$ prepared at high oxygen pressure. The electric resistivity, Seebeck coefficient and thermal conductivity of the compound show large hysteresis below the metal-insulator transition. The large value of the effective mass ($m^* \sim 8m_e$) in the metallic state indicate the narrow character of the *3d* band. The electric conduction at low temperatures (T = 2 - 20 K) is governed by the variable range hopping of the charge carriers.
Key words: Nickelates, Metal Insulator transition

## INTRODUCTION

In the perovskite oxides, rare earth nickelates RNiO$_3$ (R: rare earth), plays an important role because of the variety of physical properties and their underlying physics [1]. One of these properties is the Metal-Insulator (M-I) transitions shown by all members of RNiO$_3$ with the exception of LaNiO$_3$ which remains metallic [2,3]. The M-I transition temperature (T$_{M-I}$) depends on the size of the radius of rare earth ion 'R' and on increasing the size of rare earth ion, T$_{M-I}$ decreases from 460 K for R = Eu, to 135 K for R = Pr [4-6]. The M-I transition is accompanied by the structural transformation (from orthorhombic (Pbnm) to monoclinic (P2$_1$/m)) and magnetic ordering (paramagnetic (above T$_{M-I}$) to antiferromagnetic) at T ≤T$_{M-I}$ [6-10]. For lighter rare earth ions like Nd and Pr, antiferromagnetic ordering occurs near the T$_{M-I}$ whereas for heavier R ions this occurs below the T$_{M-I}$ [11-13].

The nature of the band gap is ascribed to different mechanisms viz. Mott Hubbard transition, charge transfer type transitions and long range charge ordering [5, 14-17]. Resonant X-ray scattering, transport and electron diffraction studies have suggested that M-I transitions results from charge ordering [18-20]. However charge ordering cannot be considered as precondition for the low temperature insulating phase for RNiO$_3$ compounds [21]. The idea of Mott-Hubbard transition which was purposed by Zaanen, Sawatzky and Allen (ZSA) in the case of late transition metal compounds, has been extended to explain the M-I transition in nickelate perovskite [22]. Based on ZSA idea, nickelates are considered as the charge transfer type insulators below the transition temperature [4,5,21,23]. The NiO$_6$ octahedra in these nickelates gets tilted slightly, that leads to the change in the electronic band structure compared to the ideal perovskite structures [4,6,24]. The tilting takes place to optimize the change in R-O bond distances introduced by the small size of rare earth ion (R). These tilting generally occur along '*b*' or '*c*' axes [6]. The amount of tilt in the NiO$_6$ octahedra is decided by the Ni-O1-Ni and Ni-O2-Ni angles, which determines the degree of overlapping of Ni-*3d* and O-*2p* orbitals and this in turn is related to electronic charge transfer [6]. The change in the Ni-O-Ni bond angle with the size of 'R' ions affects the M-I transition in nickelates [25]. The M-I transition is reported to get affected by the external pressure, and strain also [26-29].

The NdNiO$_3$ exhibits M-I transition at T ~ 205 K and orthorhombic structure at room temperature [2, 30]. The synthesis of the nickelates (with the exception of LaNiO$_3$) requires high oxygen pressure (HOP) of P ~ 0.2 – 60 kbar and high temperature of T ~ 950$^0$C [4, 26, 31, 32]. There are very few reports about the synthesis of NdNiO$_3$ at ambient oxygen pressure (AOP) [33, 34]. The electric properties of AOP phase has been studied in detail by J. Blasco *et al.* [34]. We have studied the thermal transport properties (Seebeck coefficient and thermal conductivity) along with the electrical transport properties on the clean single phase NdNiO$_3$ prepared at AOP, to bring out the differences with NdNiO$_3$ high oxygen pressure phase reported in the literature. We will call the NdNiO$_3$ prepared at ambient oxygen pressure as NdNiO$_3$ (AOP) and NdNiO$_3$ prepared at high oxygen pressure as NdNiO$_3$ (HOP), for the subsequent discussion.

## SAMPLE PREPARATION AND EXPERIMENTS

We have prepared NdNiO$_3$ at atmospheric oxygen pressure by decomposition method as used by Vassiliou *et al.* [33]. The required stoichiometric amount of the high purity Nd$_2$O$_3$ and NiO powder were homogeneously mixed by grinding and dissolved in the nitric acid (HNO$_3$). The solution was gently heated for 3 hours to remove the excess amount of nitric acid and after that obtained green color powder was slowly heated up to 400ºC. The heat treatment decomposes the mixture and the color of powder changes to black. Subsequently, the pellets were made of the obtained black powder at the pressure of around 6 kbar. The pellets were kept in the furnace at 650ºC for 5 days under the continuous flow of oxygen gas at atmosphere pressure. After the reaction, the pellets were further grounded and pelletized, and kept in the furnace for sintering at 650ºC for 48 hours. The obtained NdNiO$_3$ (AOP) showed a clean orthorhombic phase with lesser than 2 % impurity of Nd$_2$O$_3$.

The powder X-Ray diffraction (XRD) of compound was performed at room temperature using the Rigaku X-Ray Diffractometer. The electrical resistance, Seebeck coefficient, thermal conductivity and heat capacity of the compound were measured in the temperature range T = 2 – 300 K using QD-Physical Property Measurement System (PPMS) and the magnetic measurements were performed in QD-SQUID magnetometer.

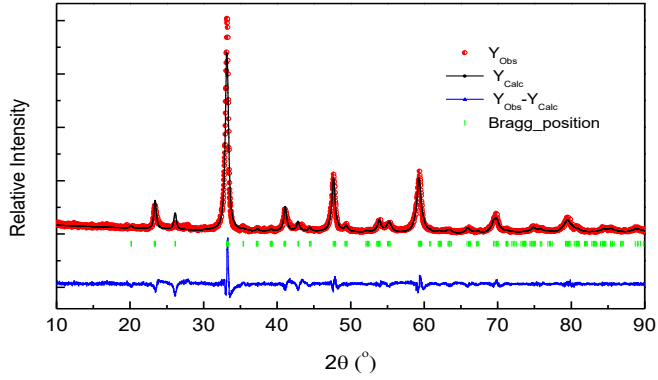

Fig. 1. X-ray powder diffraction pattern of NdNiO$_3$(AOP) obatined at room temperature.

**Table-I**

| Lattice parameters | a (Å) | b (Å) | c (Å) |
|---|---|---|---|
|  | 5.410 | 5.379 | 7.602 |
| Reliability factors | $\chi^2$ | $R_p$ | $wR_p$ |
|  | 3.43 | 0.20 | 0.26 |
| Atom | Wyckoff position | x | y | Z |
| Nd | 4c | 0.00665 | 0.96813 | 0.75 |
| Ni | 4a | 0.5 | 0 | 0.5 |
| O1 | 4c | 0.44884 | -0.04231 | 0.25 |
| O2 | 8d | 0.78470 | -0.21520 | 0.44586 |

**RESULTS**

The Fig. 1 shows the Rietveld refinement of the XRD pattern of NdNiO$_3$ compound done using the Fullprof refinement suite. The obtained NdNiO$_3$ formed in the orthorhombic unit cell (space group: Pbnm) with the lattice parameters a = 5.410Å, b = 5.379Å, and c = 7.602Å; which are in good agreement with the previous reports [4,6,16,34]. The broad nature of the XRD peaks point towards the small grain size for our sample. We estimated the average grain size of 169Å, from the width of diffraction using the Scherrer equation. We observed enhanced intensity of the *(2 0 0)* plane (at *2θ ~ 33.2⁰*) and the suppression of the peak for *(1 1 1)* plane (at *2θ ~ 26.1⁰*), which point towards the preferred orientation along the *(1 1 1)* plane. Such stacking faults indicates the laminar disorder in our sample, which may be related to the variation in the Oxygen stoichiometry. The refined lattice parameters, reliability factors and atomic positions, are given in the table - 1.

The estimated values of the bond angle Ni-O-Ni, corresponding to both the oxygen atoms (O1 and O2) are 157.19(4)°, and 156.99(2)° for Ni-O1-Ni and Ni-O2-Ni respectively. The average value of the Ni-O-Ni angle (157.09°) is comparable to that for the other nicklates RNiO$_3$; viz. LaNiO$_3$ (165.2°), PrNiO$_3$ (158.1°), NdNiO$_3$ (157° to 158°), and SmNiO$_3$ (152.6°) [6]. The tolerance factor (t = $d_{Nd-O}/\sqrt{2} d_{Ni-O}$) of 0.987 was obtained for the compound from the average bond distances between Ni-O and Nd-O atoms, which is in agreement with the reports on NdNiO$_3$ (AOP) but higher than NdNiO$_3$ (HOP) [4,6].

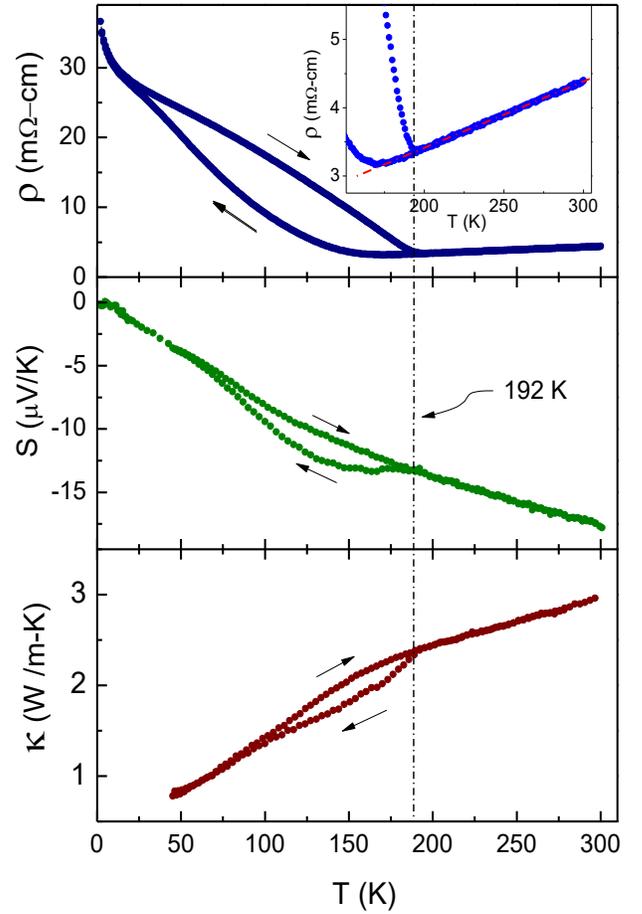

Fig.2. Temperature dependence of the electrical resistivity, ρ(T), Seebeck Coefficient, S(T), and Thermal conductivity (κ) of the NdNiO$_3$(AOP).

The electrical resistivity, Seebeck coefficient, and thermal conductivity of the compound are shown in the fig. 2. There is clear transition from the metallic to insulating phase at T ~ 192 K, which is lower than the metal-insulator transition temperature (T$_{M-I}$) of 205 K for the NdNiO$_3$ (HOP) [30]. The electrical resistivity ρ(T) of the compound increases to 37 mΩ-cm at 2 K from 3.4 mΩ-cm at T$_{M-I}$. The rise in ρ(T) in NdNiO$_3$ (AOP) is smaller compared to NdNiO$_3$ (HOP), where ρ(T) increases up to 5 order of magnitude in the similar temperature range [35]. The room temperature resistivity ρ$_{(T=300\,K)}$ ~ 4.38 mΩ-cm is 3.6 times greater than the 1.2 mΩ-cm for NdNiO$_3$ (HOP) [30]. However the normalized resistivity slope (1/R(dR/dT ~ 3.82×10$^{-3}$ K$^{-1}$) is of same order as for the good metals (e.g. 3.8 × 10$^{-3}$ K$^{-1}$ for Ag at T = 293 K) [36].

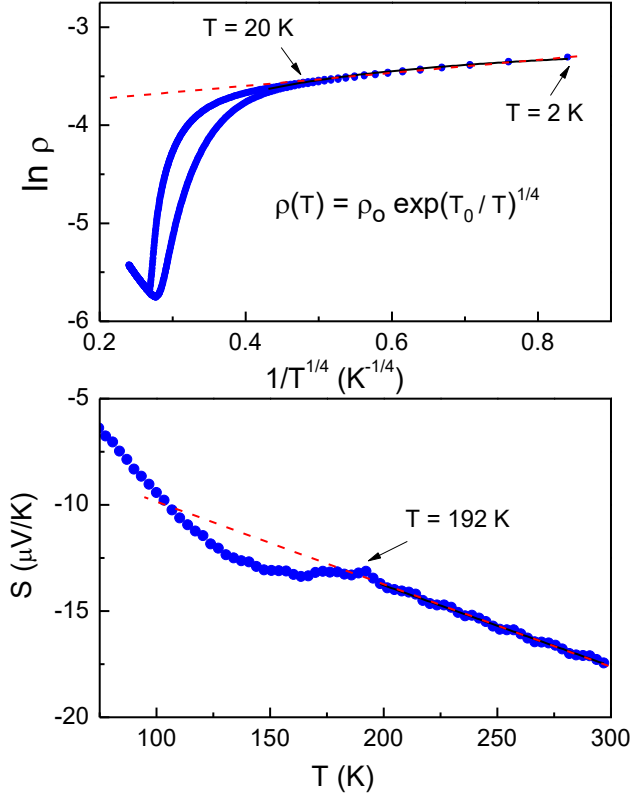

Fig. 3. The *ln ρ* versus *1/T$^{1/4}$* curve (upper panel) showing conduction by variable range hopping for T ≤ 20 K. The temperature dependence of Seebeck coefficient S (lower panel) shows linear dependence with T in the metallic regime.

The hysteretic behavior shown in resistivity curve ρ(T), extends down to 20 K from T$_{M-I}$, with hysteresis width ΔT$_H$ ~170 K, which is much larger than (ΔT$_H$ ~ 85 K) for the NdNiO$_3$ (HOP) [9, 30].

The ln ρ versus 1/T$^{1/4}$ plot in fig. 3 shows that conduction mechanism is governed by the Variable Range Hopping (VRH) in the range 2 K ≤ T ≤ 20 K. The resistivity follows ρ(T) = ρ$_0$exp(T$_0$/T)$^{1/4}$ behavior for the disorder system where conduction takes place by the hopping of charge carriers to the vacant neighboring and next neighbor sites. It is possible that the nickelate phase prepared at ambient oxygen pressure consist a number of oxygen vacancies and disorder and thus facilitating the conduction by variable range hopping.

The Seebeck coefficient (S) also show anomaly (shown in fig. 2) at the M-I transition with the clear hysteresis. The room temperature value of S ~ - 18.0 μV/K match well with the reported values [30, 37]. The magnitude of the Seebeck coefficient |(S(T)|) decreases at T < T$_{M-I}$ depicting the presence of enough charge carriers below the transition. The negative Seebeck coefficient suggests that dominant charge carriers are electrons. With the decrease of temperature, the |S(T)| decreases in the whole temperature range. On cooling from room temperature, |S(T)| shows slight increase (plateau like behavior) below T$_{M-I}$, before decreasing again down to 2 K. This increase in |S| at T$_{M-I}$ is very small compared to that observed for the NdNiO$_3$ (HOP) but their room temperature values are quite comparable [30]. Under the relaxation time approximation, the Seebeck coefficient of the metals can be given by the Drude's expression [38];

$$S(T) = -\frac{\pi^2 k_B}{3e}\frac{T}{E_F}\left\{\frac{g(\varepsilon)}{n} + \frac{\partial \ln[\tau(\varepsilon)]}{\partial \varepsilon}\right\}_{\varepsilon=E_F} \quad \ldots\ldots\ldots (1)$$

where g(ε) is density of states, n is the charge carrier density and E$_F$ is the Fermi energy. τ(ε) is the energy dependent relaxation time generally written as τ(ε) ~ ε$^\alpha$ where the exponent α is related to scattering mechanism. For particular case of α = -1 i.e. τ(ε) ~ 1/ε; under the parabolic band approximation, Drude's expression in its simplest form can be given as below;

$$S(T) = -\frac{\pi^2 k_B}{6e}\frac{T}{E_F}\ldots\ldots\ldots (2)$$

The linear temperature fit of the data in the metallic region (fig. 3) gives the slope *dS(T)/dT* ≈ -0.03888 μV/K$^2$. Using this slope value in equation (2) we obtained the Fermi energy *E$_F$* ≈ 0.32 eV which is smaller than the *E$_F$* ≈ 0.41 eV for NdNiO$_3$ (HOP), *E$_F$* ≈ 0.40 eV for PrNiO$_3$, and *E$_F$* ≈ 0.21 eV for LaNiO$_3$ [30, 39, 40].

The κ(T) of the compound decreases below T$_{M-I}$ with a clear hysteresis. The loss of charge carriers at T$_{M-I}$ may lead to the decrease in the value of κ(T) at T$_{M-I}$. Previous reports have suggested that the metallic phase in PrNiO$_3$ and NdNiO$_3$ does not disappear completely for T < T$_{M-I}$, and the oxygen vibrations play an important role in T$_{M-I}$ [34,37,41]. These oxygen vibrations might play an important role in the thermal conductivity of the compound that can be further enhanced in the NdNiO$_3$ (AOP). In the absence of the low temperature data below 40 K, we could not make estimation of the phononic (κ$_l$) and electronic (κ$_e$) contribution to the total thermal conductivity.

The heat capacity (HC) data for NdNiO$_3$ (AOP) in the temperature range 2 - 260 K is shown fig. 4. We did not observe any peak like anomaly near M-I transition. However a deflection in the heat capacity (HC) near around $T_{M-I}$ can be clearly observed in the $C_p/T$ versus T plot (shown in the inset of fig. 4). Our data matches well with the report for the samples prepared at ambient pressure condition [42]. Also there is no change in specific heat in the presence of magnetic field of up to H = 5 T (data not shown here). We observed a slight increase in the HC value for T ≤ 10 K, which is difficult to understand at this point. Therefore it is not possible to extract the electronic ($\gamma$) and phononic contribution ($\beta$) from the $C_p(T) = \gamma T + \beta T^3$ from the $C_p/T$ versus $T^2$ curve. However the estimated value of $\gamma$ (11.24 mJ/mol-K$^2$) from $E_F$ obtained from the Seebeck coefficient data is close to the reported values of 8.2 mJ/mol-K$^2$ for NdNiO$_3$ (HOP) [30].

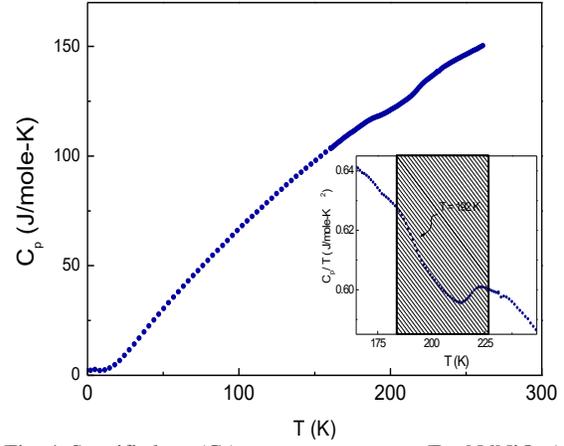

Fig. 4. Specific heat (C$_p$) versus temperature For NdNiO$_3$ (AOP). The C$_p$/T vs T curve in the inset show the deviation in specific heat at M-I transition.

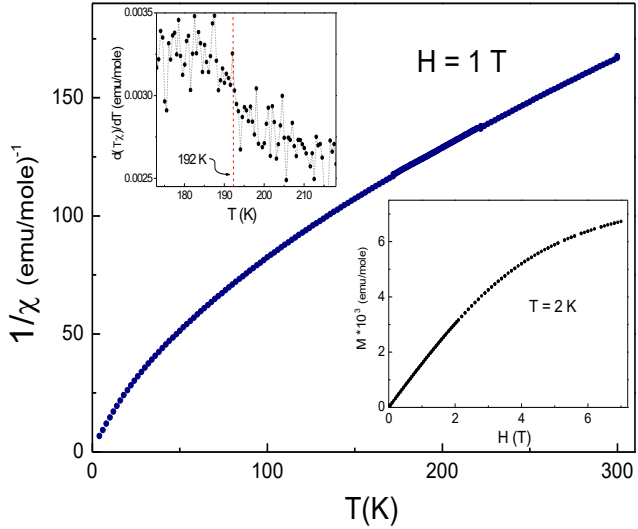

Fig. 5. The inverse magnetic susceptibility ($1/\chi$) versus temperature measured at H = 1 Tesla. M versus H data taken at T = 2 K is shown in the lower inset. The $d(T\chi)/dT$ versus T plot in upper inset shows an anomaly at M-I transition.

The dc magnetic susceptibility ($\chi(T)$) data taken at H = 1 Tesla is plotted as $1/\chi(T)$ versus T in the fig. 5. We did not observe any peak in the susceptibility data at the Neel temperature $T_N$. It has been reported for the NdNiO$_3$ (HOP) phase that M-I transition is accompanied by the paramagnetic to antiferromagnetic transition with the Neel temperature ($T_N$) close to the $T_{M-I}$ [21]. We have plotted $d(T\chi)/dT$ versus T around the $T_{M-I}$ (shown in the inset of the fig. 5), to extract the information about the antiferromagnetic (AFM) phase transformation near $T_{M-I}$. There is small anomaly in the $d(T\chi)/dT$ versus T around $T_{M-I}$ = 192 K. The M-H data taken at T = 2 K is shown in the right inset of the fig. 5, where a saturation trend in magnetization is apparent at higher magnetic field which is an indication of the presence of the ferromagnetic exchange term. The ferromagnetic exchange term might arise due to small level impurities of ferromagnetic Ni atom or due to oxygen defects in the compounds. Though, we did not observe peak corresponding to the Ni, but our XRD data is not sensitive for < 1% impurities so the possibility of a smaller amount of Ni impurity cannot be denied. The magnetization data has been fitted using Curie-Weiss equation $\chi = \chi_o + C/(T - \theta_{CW})$; where C is the Curie constant, $\theta_{CW}$ is the Curie-Weiss temperature and $\chi_o$ is the temperature independent susceptibility. The magnetization data do not fit well to the Curie-Weiss equation in the measured temperature range of 2- 300 K. The crystal field splitting of the Nd$^{3+}$ (J = 9/2 ground state) at lower temperature may give rise to the change in the van Vleck susceptibility contribution to $\chi_o$ and hence bringing the change in the Curie constant value with the temperature variation. Therefore we have tried to fit our data in different temperature regimes. The obtained values of the fitting parameters are shown in table-II. Considering the fitting range to be 15 - 300 K, we find $\chi_o = 3.36 \times 10^{-3}$ emu/mole, and $\mu_{eff}$ = 2.68 $\mu_B$ per formula unit. Though the $\mu_{eff}$ for the compound is less compared to the expected theoretical values due to the magnetic Nd$^{3+}$ ion (3.62 $\mu_B$), and Ni$^{3+}$ (0.9 $\mu_B$) ions, it is comparable to the $\mu_{eff}$ value reported by Vassiliou *et al.* for the NdNiO$_3$ (AOP) phase [21, 33]. The fitting of the magnetization data in the paramagnetic region above the AFM transition in the temperature range 200-300 K, we obtained $\chi_o$ = 1.64 × 10$^{-3}$ emu/mole, and $\mu_{eff}$ = 3.39 $\mu_B$ per formula unit, which are comparable to $\chi_o$ = 0.97 × 10$^{-3}$ emu/mole, and $\mu_{eff}$ = 3.57 $\mu_B$ per formula unit for the NdNiO$_3$ (AOP) reported by Blasco *et al.* [34].

**Table-II**

| T (K) | $\chi_o$ (emu/mole) | $\theta_{CW}$ (K) | C (emu K/mole) | $\mu_{eff}$ ($\mu_B$ per f.u.) |
|---|---|---|---|---|
| 2 – 300 | 4.41×10$^{-3}$ | - 3.2 | 0.768 | 2.48 |
| 15 – 300 | 3.36 ×10$^{-3}$ | -5.8 | 0.904 | 2.68 |
| **200 - 300** | **1.64×10$^{-3}$** | **-34.0** | **1.44** | **3.39** |

## 4. DISCUSSION

The NdNiO$_3$ (AOP) shows similar electronic transport properties above $T_{M-I}$ as for NdNiO$_3$ (HOP). However there is significant difference in the insulating phases of these two compounds prepared at the two different oxygen pressures. In contrast to the several order of increase in the magnitude of resistivity after $T_{M-I}$ in high oxygen pressure compounds, the ambient oxygen pressure phase is less insulating and resistivity increases only one order of magnitude. The electrical and thermal transport properties of the compound around the M-I transition depends on the percentage volume fraction of the metallic and insulating phases and their transformation dynamics. According to Granados *et al.,* these materials have a mixture of metallic and insulating phase with different transport characteristics [30, 39]. Our resistivity and thermal transport data show a broad hysteresis extending from $T_{M-I}$ (~192 K) to 20 K with hysteresis width of $\Delta T_H \sim$ 170 K. The insulating phase in our compound is in incipient stage and does not grow like the NdNiO$_3$ (HOP). The room temperature (300 K) value of resistivity for NdNiO$_3$ (AOP) is 4.3 mΩ-cm, which is approximately 3.5 more than that for NdNiO$_3$ (HOP) [30].

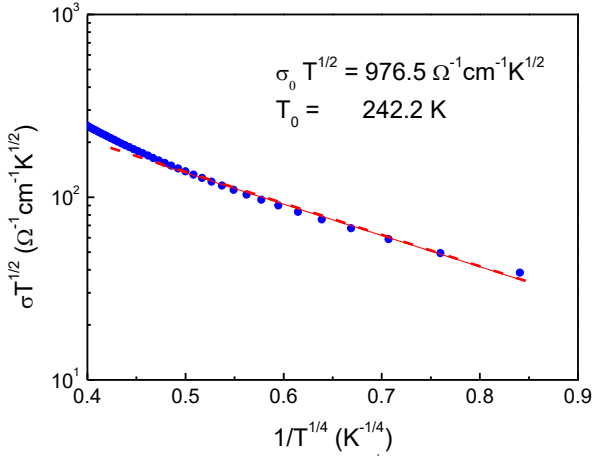

Fig. 6. The semi-logarithmic plot of $\sigma\sqrt{T}$ versus $1/T^{1/4}$ plot.

The low temperature (T ≤ 20 K) resistivity of the compound suggests the conduction by the Variable Range Hopping. The presence of charge disorder and oxygen vacancy along with the high value of charge carrier facilitates the hopping of the carriers to neighboring and next neighboring sites. We have calculated $k_Fl$ for the compound ($k_F$ is Fermi wave vector and $l$ is mean free path) using the carrier density $n$ (obtained from the refined crystal structure data) and resistivity values from the experimental data. The $k_Fl$ values at 2 K and $T_{M-I}$ (~192 K) are 0.041 and 0.44 which are very low compared to Ioffe Regel criterion of $k_Fl \sim 1$ for M-I transition suggesting the built in disorder in the compound. The low temperature (T < 20 K) electrical conductivity follows the Mott's expression $\sigma = \sigma_0\,exp[-(T_0/T)^{1/4}]$ for conductivity by the hopping of carriers to distant sites via localized states near Fermi energy. In the derivation of this expression, Mott used the parameters; $\alpha$, the inverse rate of fall off of the electron wave function associated with the localized state near Fermi energy, and $N(E_F)$, the density of localized states at the Fermi Energy [43 - 45]. The rigorous studies on Mott's expression, considering the various conditions of temperature, field and distribution of hopping showed the constants $\sigma_0$ and $T_0$ to have functional form; $\sigma_0 = e^2 a^2 \nu_{ph} N(E_F)$, and $T_0 = \lambda \alpha^3 / k_B N(E_F)$, where $e$ is the electronic charge, $a$ is the hopping distance, $\nu_{ph}$ is the phonon frequency (~$10^{13}$ sec$^{-1}$), $k_B$ is Boltzmann's constant, $\lambda$ is a dimensionless constant (~18.1), and $N(E_F)$ is the density of states at Fermi energy [43-45]. The hopping distance '$a$', is given by the expression; $a = [9/8\pi k_B \alpha T N(E_F)]^{1/4}$, which shows that $\sigma_0$ has $1/\sqrt{T}$ dependence on the temperature [44]. We have plotted $\sigma\sqrt{T}$ versus $1/T^{1/4}$ in the fig. 6 for the temperature range 40 K ≤ T ≤ 2 K. We obtained a liner fit to the curve below 20 K, which is smaller than the range 60 K ≤ T ≤ 4.2 K reported by the Blasco et al. [34]. The linear fit to the curve gives $\sigma_0\sqrt{T}$ and $T_0$ as 9.77×10$^2$ Ω$^{-1}$cm$^{-1}$K$^{1/2}$, and 242.2 K respectively. These values are smaller than the values reported by Blasco et al. [34]. However, If we take the linear fitting to their low temperature data in the similar temperature range as ours, these parameters come out to be 6.71×10$^2$ Ω$^{-1}$cm$^{-1}$K$^{1/2}$, and 270 K, which are comparable to our values. The density of states at Fermi energy $N(E_F)$ can be obtained from $\sigma_0\sqrt{T}$ and $T_0$ values using the expression [45];

$$N(E_F) = \left(1.996 \times 10^{48}/\nu_{ph}^3\right) \times \left[\left(\sigma_0\sqrt{T}\right)^3 \sqrt{T_0}\right] \text{cm}^{-3}\,\text{eV}^{-1}$$

……….. (3)

We obtained a density of states $N(E_F)$ in the low temperature regime as 2.89×10$^{19}$ eV$^{-1}$ cm$^{-3}$, which is comparable to the recalculated values of ~ 1.0 ×10$^{19}$ eV$^{-1}$cm$^{-3}$ from the low temperature data of Blasco et. al. However Blasco et al. have reported $N(E_F)$ value as 2.3 ×10$^{20}$ eV$^{-1}$cm$^{-3}$ from the fitting range of 60 K ≤ T ≤ 4.2 K, which is 8 times larger than the value obtained by us. The hopping distance or the average polaron radius 217 Å (obtained from the $N(E_F)$ and $\alpha$ value of 3.224×10$^5$ cm$^{-1}$), is much higher compared to the lattice size, and indicates towards the large polaron.

For T > $T_{M-I}$, the $\rho(T)$, and the $|S(T)|$ increases linearly with temperature similar to the metallic system. The Fermi energy ($E_F \sim 0.315$ eV) in the metallic state obtained under the assumption of one electron per Ni atom (Ni$^{3+}$, $t_{2g}^6$ $e_g^1$), for the cell volume V = 221.22 Å$^3$, and carrier density, $n = 1.808\times 10^{22}$/cm$^3$ gives the density of states $g(\varepsilon) = 8.62\times 10^{22}$ states/eV-cm$^3$ (from the relation, $g(\varepsilon) = 3n/2E_F$), which is comparable to the reported values [30]. Using bare density of states $g^o(\varepsilon)$ corresponding to free electrons and $g(\varepsilon)$, we obtained the effective mass $m^* = 8m_e$. This value is higher than $6m_e$ predicted by Granados et al. for the NdNiO$_3$ (HOP) [30]. Higher values of $g(\varepsilon)$ and $m^*$ for NdNiO$_3$(AOP) suggests the metallic nature above the $T_{M-I}$ under the assumption of one electron per Ni atom. In the Brinkman-Rice model for correlated electrons, effective mass ($m^*$) is related to the correlation energy $U$ by the expression [46]; $m^*/m_e = 1/[1 - (U/U_0)^2]$, where $U_0$ is the energy at which the correlation gap opens. The effective mas of $m^* = 8m_e$, gives the correlation energy $U \sim 0.9U_0$, which indicate the system at the verge of M-I transition. For the charge transfer type band gap in the insulating state, and metallic conductivity via overlapping of oxygen $2p$ and Nickel $3d$ bands, the enhanced value of the effective mass points towards the very narrow character of the $3d$ band [5].

The Ni-O-Ni bond angle for NdNiO$_3$(AOP) is comparable to high oxygen pressure phase of NdNiO$_3$ and PrNiO$_3$ but lower than that for the LaNiO$_3$. The larger bond angle leads to the larger bandwidth and hence to the higher metallicity in the system. The higher value of the tolerance factor t ~ 0.987 for the compounds NdNiO$_3$ (AOP), in comparison to that for NdNiO$_3$ (HOP), indicate lesser distortion (tilting) in the NiO$_6$ octahedra, and contribute to the metallic component in the compound.

5. **CONCLUSION**

The electronic and thermal transport properties of NdNiO$_3$ (AOP) are compared with the NdNiO$_3$ (HOP). The electrical resistivity, Seebeck coefficient and the thermal conductivity of the NdNiO$_3$ (AOP) compounds show anomaly at the metal-insulator transition with the broad hysteresis. In the metallic regime, the electronic properties of NdNiO$_3$ (AOP) can be described in term of the Fermi gas of electrons of high effective mass (~ 8$m_e$). The NdNiO$_3$ (AOP) shows lower value of resistivity below T$_{M-I}$ compared to NdNiO$_3$ (HOP). The resistivity of the compound below T$_{M-I}$ is governed by the localization of charge carriers aided by the disorder, and conduction takes place by the variable range hopping of the charge carriers for $2 \leq T \leq 20$ K. The laminar defects, oxygen vacancies and the smaller grain size might be playing an important role in the low temperature physical properties of the NdNiO$_3$ (AOP). Further experiments like photoelectron spectroscopy (PES), and inelastic x-ray scattering for ascertaining the density of states at fermi energy and the relative fraction of the Ni$^{2+}$ and Ni$^{3+}$ occupation of Ni $3d$ band, respectively; would help in discerning their role in the metal-insulator transition in the less insulating NdNiO$_3$ (AOP) and more insulating NdNiO$_3$ (HOP) phases.

**ACKNOWLEDGEMENT:** CSY acknowledges the seed grant No. IITMandi/SG/ASCY/29, for financial support and Advanced Material Research Center (AMRC), IIT Mandi for the experimental facilities used for this study. MKH acknowledges the IIT Mandi for the HTRA fellowship.

**REFERENCES:**

[1] J.B.Goodenough, Prog. Sol. State Chem. 5 (1972) 145.
[2] M.L.Medarde, J. Phys.: Condens. Matter 9 (1997) 1679.
[3] J.Son, Ph.D. thesis, University of California, Santa Barbara (2011).
[4] P.Lacorre, J.B.Torrance, J.Pannetier, A.I.Nazzal, P.W.Wang, T.C.Huang, J. Solid State Chem. 91 (1991) 225.
[5] J.B.Torrance, P.Lacorre, A.I.Nazzal, E.J.Ansaldo, Ch.Niedermayer, Phys. Rev. B 45 (1992) 8209.
[6] J.L.Garcia-Munoz, J.Rodriguez-Carvajal, P.Lacorre, J. B.Torrance, Phys. Rev. B 46 (1992) 4414.
[7] J.A.Alonso, M.J.Martinez-Lope, M.T.Casais, M.G A.Aranda, M.Fernandez-Diaz, J. Am. Chem. Soc. 121 (1999) 4754.
[8] J.S.Zhou, J.B.Goodenough, B.Dabrowski, Phys. Rev. Lett. 95 (2005) 127204.
[9] J.L.Garcia-Munoz, M.G.A.Aranda, J.A.Alonso, M.J.Martinez-Lope, Phys. Rev. B 79 (2009) 134432.
[10] A.J.Hauser, Mikheeve, N.E.Moreno, T.A.Cain, J.Hwang, J.Y.Zhang, S.Stemmer, Appl. Phys. Lett. 103 (2013) 182105.
[11] J.S.Zhou, J.B.Goodenough, B.Dabrowski, Phys. Rev. Lett. 94 (2005) 226602.
[12] J.L.Garcia-Munoz, M.Amboage, M.Hanfland, J.A.Alonso, M.J.Martinez-Lope, R.Mortimer, Phys. Rev. B 69 (2004) 094106.
[13] M.L.Medarde, A.Fontaine, J.L.Garcia-Munoz, J.Rodriguez-Carvajal, M.DeSantis, M.Sacchi, G.Rossi, P.Lacorre, Phys. Rev. B 46 (1992) 975.
[14] M.K.Stewart, J.Liu, M.Kareev, J.Chakhalian, D.N.Basov, Phys. Rev. Lett. 107 (2011) 76401.
[15] U.Staub, G.I.Meijer, F.Fauth, R.Allenspach, J.G.Bednorz, J.Karpinski, S.M.Kazakov, L.Paolasini, F.D'Acapito, Phys. Rev. Lett. 88 (2002) 126402.
[16] V.Scagnoli, U.Staub, M.Janousch, A.M.Mulders, M.Shi, G.I.Meijer, S.Rosenkranz, S.B.Wilkins, L.Paolasini, J.Karpinski, S.M.Kazakov, S.W.Lovesey, Phys. Rev. B 72 (2005) 155111.
[17] J.S.Zhou, J.B.Goodenough, B.Dabrowski, P.W.Klamut, Z.Bukowski, Phys. Rev. B 61 (2000) 4401.
[18] M.Zaghrioui, A.Bulou, P.Lacorre, P.Laffez, Phys. Rev. B 64 (2001) 081102.
[19] M.H.Upton, Yongseong Choi, Hyowon Park, J.Liu, D.Meyers, J.Chakhalian, S.Middey, Jong-Woo Kim, J.R.Philip, Phys. Rev. Lett. 115 (2015) 036401.
[20] J.Zaanen, G.A.Sawatzky, J.W.Allen, Phys. Rev. Lett. 55 (1985) 418.
[21] J.L.Garcia-Munoz, J.Rodriguez-Carvajal, P.Lacorre, Europhys. Lett. 20 (1992) 241.
[22] I.I.Mazin, D.I.Khomskii, R.Lengsdorf, J.A.Alonso, W.G.Marshall, R.M.Ibberson, A.Podlesnyak, M.J.Martinez-Lope, M.M.Abd-Elmeguid, Phys. Rev. Lett. 98 (2007) 176406.
[23] M.Imada, A.Fujimori, Y.Tokura, Rev. Mod. Phys. 70 (1998) 1039.
[24] X.Obradors, L.M.Paulius, M.B.Maple, J.B.Torrance, A.I.Nazzal, J.Fontcuberta, X.Granados, Phys. Rev. B 47 (1993) 12353.
[25] M.Medarde, JMesot, S.Rosenkranz, P.Lacorre, W.Marshall, S. Klotz, J. S. Loveday, G. Hamel, S. Hull, P. Radaelli, Physica B 234-236 (1997) 15.
[26] P.C.Canfield, J.D.Thompson, S.W.Cheong, L.W.Rupp, Phys. Rev. B 47 (1993) 12357.
[27] A.Tiwari, C.Jin, J.Narayan, Appl. Phys. Lett. 80 (2002) 4039.
[28] J.L. Garcia-Munoz, J.Rodriguez-Carvajal, P.Lacorre, Phys. Rev. B 50 (1994) 978.
[29] V.Scagnoli, U.Staub, Y.Bodenthin, M.Garcia-Fernandez, A.M.Mulders, G.I.Meijer, G.Hammerl, Phys. Rev. B 77 (2008) 115138.
[30] X.Granados, J.Fontcuberta, X.Obradors, Phys. Rev. B 48 (1993) 11666.
[31] G.Demazeau, A.Marbeuf, M.Pouchard, P.Hagenmuller, J. Solid State Chem. 3 (1971) 582.
[32] N.E.Massa, J.A.Alonso, M.J.Martinez-Lope, I.Rasines, Phys. Rev. B 56 (1997) 986.
[33] J.K.Vassiliou, M.Hornbostel, R.Ziebarth, F.J.Disalvo, J. Solid State Chem. 81 (1989) 208.
[34] J.Blasco, M.Castro, J.Garcia, J.Phys.:Condens. Matter 6 (1994) 5875.
[35] Devendra Kumar, K.P.Rajeev, J.A.Alonso, M.J.Martinez-Lope, J. Phys.: Condens. Matter 21 (2009) 485402.
[36] R.A.Serway, Principles of Physics (2$^{nd}$ed.)(1998).
[37] V.B.Barbeta, R.F.Jardim, M.T.Escote, N.R.Dilley, J. Appl. Phys. 101 (2007) 09N509.
[38] N.W.Ashcroft, N.D.Mermin, Solid State Physics (1976).
[39] X.Granados, J.Fontcuberta, X.Obradors, J.B.Torrance, Phys. Rev. B 46 (1992) 15683.
[40] J.P.Kemp, P.A.Cox, Solid-State Commun. 75 (1990) 731.
[41] J.S.Zhou, J.B.Goodenough, B.Dabrowski, Phys. Rev. B 67 (2003) 020404.
[42] V.B.Barbeta, R.F.Jardim, M.S.Torikachvili, M.T.Escote, F.Cordero, F.M.Pontes, F.Trequattrini, J. Appl. Phys. 109 (2011) 07E115.
[43] D.K. Paul, S.S. Mitra, Phys. Rev. Lett. 31 (1973) 1000.
[44] V. Ambegaokar, B.I. Halperin, J.S. Langer, Phys. Rev. B 4 (1971) 2612.
[45] A. Lewis, Phys. Rev. Lett. 29 (1972) 1555.
[46] W.F. Brinkman and T.M. Rice, Phys. Rev. B 2 (1970) 4302